# An algorithm for detecting oscillatory behavior in discretized data: the damped-oscillator oscillator detector

David Hsu, Murielle Hsu, He Huang and Erwin B. Montgomery, Jr

Department of Neurology

University of Wisconsin Hospital & Clinics

Madison WI 53792

Email: hsu@neurology.wisc.edu

## Abstract

We present a simple algorithm for detecting oscillatory behavior in discrete data. The data is used as an input driving force acting on a set of simulated damped oscillators. By monitoring the energy of the simulated oscillators, we can detect oscillatory behavior in data. In application to in vivo deep brain basal ganglia recordings, we found sharp peaks in the spectrum at 20 and 70 Hz. The algorithm is also compared to the conventional fast Fourier transform and circular statistics techniques using computer generated model data, and is found to be comparable to or better than fast Fourier transform in test cases. Circular statistics performed poorly in our tests.

## Introduction

Oscillatory behavior underlies many physical and biological phenomena, including the electrical behavior of neurons and neural systems. Data from neural recordings are often saved in the form of a list of discrete times when a given neuron fires an action potential.

Such discretized data often appear randomly distributed in time, i.e., the spiking behavior is often well-described as a Poisson process for which the distribution of interspike intervals then is exponential. However, sometimes these data contain hidden oscillations, not obvious to casual inspection. These oscillations can be revealed by standard procedures, such as the Fast Fourier Transform (FFT) and circular statistics (CS), or by more advanced procedures including the multi-taper method [1-3].

These methods all have their strengths and weaknesses. Here we propose a simple alternative. In our approach, the experimental data are employed as an input driving force acting on a set of lightly damped computer-simulated harmonic oscillators. These oscillators we refer to as the "mathematical" oscillators. If there is a hidden oscillator in the data (a "data oscillator"), then it will drive a mathematical oscillator in a resonant way if their frequencies are the same. By monitoring the time rate of change of the energy of each mathematical oscillator, we can then detect when a data oscillator appears and disappears. The appearance of a data oscillator will cause a change in the energy of the mathematical oscillator of the same frequency, and the disappearance of that data oscillator will cause the energy of the corresponding mathematical oscillator to stop changing. Furthermore, if we lightly damp the mathematical oscillator, then we can inhibit artifactual responses of the mathematical oscillator.

We refer to our method as the damped-oscillator oscillator detector (DOOD). When tested on a time series recorded from the basal ganglia of a non-human primate, the DOOD algorithm picked out oscillators at 20 and 70 Hz, while neither FFT nor CS was able to do so. We also tested the DOOD algorithm on computer generated data, to simulate a 70 Hz periodic process in the presence of Poisson noise. We found the DOOD algorithm to be superior to FFT, while FFT was superior to CS.

## Methods

The data $h(t)$ consist of either a continuous stream of values at every time instant $t$ with time increment $dt$, or else it may consist of simply a list of times when discrete events occur. We consider the latter case for illustration. The experimental data is derived from electric potential recordings acquired from electrodes inserted into the basal ganglia of non-human primates. Whenever a unit neuronal action potential is identified, the time of discharge is automatically recorded with a precision of 0.04 msec, along with a code identifying each individual neuronal unit. The amplitude of the action potential is not recorded, nor is the duration. Thus we represent neuronal discharges by taking $h(t) = h_0 = 1$ for a duration of 1 msec for each neuronal discharge, and $h(t) = 0$ otherwise.

Turning our attention to the mathematical oscillators, the $n^{th}$ mathematical oscillator has frequency f(n) = n (in Hz), with n = 1 to N and mass taken to be unity in arbitrary units. Neglecting for now the index for n, the equation of motion for the $n^{th}$ mathematical oscillator is then

$$\ddot{x}(t) = h(t) - \omega_0^2 x(t) - 2g\dot{x}(t) \qquad \text{Eq (1)}$$

Here $\omega_0$ is the natural frequency of oscillation for this oscillator, in the absence of friction, while $g$ is the friction constant. We break the data down into time windows small enough such that $h(t)$ is a constant within each time window. Consider the time window $[t_a, t_b)$. The solution for Eq (1), within this time window, is then

$$x(t) = \frac{h}{\omega_0^2} + \exp(-gt)[a\cos(\omega t) + b\sin(\omega t)] \quad \text{Eq (2)}$$

where

$$\omega^2 = \omega_0^2 - g^2 \quad \text{Eq (3)}$$

and

$$\omega = 2\pi f\,. \quad \text{Eq (4)}$$

Here $f$ is the frequency of the n[th] mathematical oscillator, after accounting for friction, and $a$ and $b$ are constants which are chosen so that $x(t)$ and $\dot{x}(t)$ are continuous across neighboring time windows. The frequency $\omega_0$ can be related to $f$ through Eqs (3) and (4).

The friction constant $g$ dampens oscillations after a time period on the order of $1/g$. We choose $g$ such that the timescale of frictional damping is in terms of the inherent period of the mathematical oscillator:

$$g = g_0\omega. \quad \text{Eq (5)}$$

This choice allows us to choose the same value of $g_0$ for mathematical oscillators of different frequencies. Unless the data oscillator continues to drive the mathematical oscillator often enough, relative to the natural cycling period of the damped oscillator, then the oscillation will damp out. Thus we can use $g_0$ to help remove artifactual peaks in the power spectrum. That is, if the driving force $h(t)$ does not consistently and persistently drive the mathematical oscillator, then even a small frictional term will damp out that oscillator. On the other hand, if $h(t)$ does consistently and persistently drive the mathematical oscillator, then the driving term overcomes the friction term as long as the friction is not too large. In practice, we will start with a frictionless set of mathematical oscillators, observe how many peaks there are in the power spectrum of these oscillators, then turn up the friction slowly and see which peaks remain. Those oscillators that persist as the friction is turned up are more likely to be real and not artifactual.

The instantaneous energy of the mathematical oscillator is

$$E = \frac{1}{2}\left(\dot{x}^2 + \omega_0^2 x^2\right). \qquad \text{Eq (6)}$$

Taking the time derivative of Eq (6) and using Eq (1), the time rate of change of the energy $\dot{E} \equiv S$ can then be written

$$S = \dot{x}(h - 2g\dot{x}). \qquad \text{Eq (7)}$$

The procedure for initiating and running the algorithm is now as follows: (1) start with the mathematical oscillator at rest ($x = 0$ and $\dot{x} = 0$, $a = 0$, $b = 0$), (2) set the first time window of interest to be $[t_a, t_b)$, with $t_a = 0$ and $t_b$ = time of the first neuronal discharge, (3) this time window is a trivial time window in which the oscillator remains at rest, (4) define a new time window of interest to be an event window $[t_a(new), t_b(new))$, with the new $t_a$ being equal to the prior $t_b$, and the new $t_b(new) = t_a(new) + 1$ msec, (5) solve for the new $a$ and $b$ in this time window given that $x$ and $\dot{x}$ are continuous across the boundary of the current and the prior time windows, (6) propagate Eq 2 out to time $t_b$, (7) reset the time window of interest such that the new $t_a$ is the old $t_b$, and the new $t_b$ is the time of the next neuronal discharge, (8) this time window is now one of free propagation, (9) propagate Eq 2 in the current time window of interest out to $t_b$ with $h = 0$, (10) return to step (4). This process is continued until the end of the data set, while either monitoring the running time average of $S(t)$ or saving cumulative values of $S(t)$ for a calculation of the mean $S(t)$ at the end. The entire process is then repeated for the next mathematical oscillator with a different frequency $f$.

A convenient way of calculating a weighted running average of *S(t)* is as follows. Let $\tau$ be the time duration of the most recent free propagation (time window in step 8 above) and let $\tau_0$ be the timescale over which one wishes to average. Then define:

$$S_a(t) = S(t) + \exp(-\tau/\tau_0) S_a(t - dt) \qquad \text{Eq (8)}$$

$$S_b(t) = 1 + \exp(-\tau/\tau_0) S_b(t - dt) \qquad \text{Eq (9)}$$

Then $S_0(t) = S_a(t) / S_b(t)$ gives a running average of *S(t)* such that the most recent values of *S(t)* are weighted more heavily. A sample Fortran program is available on request.

A further refinement is possible. If we choose our set of mathematical oscillators to be of frequency n = 1 to 500 Hz in increments of $\Delta f = 1$ Hz, then we intend that each mathematical oscillator represent all data oscillators within 0.5 Hz of its own resonant frequency. That is, a mathematical oscillator at 20 Hz is meant to detect all data oscillators from 19.5 to 20.5 Hz. To allow a mathematical oscillator of frequency *f* to respond to data oscillators with natural cycling periods between $T_{min} = 1/(f + \Delta f/2)$ and $T_{max} = 1/(f - \Delta f/2)$, we randomize neuronal discharge times by adding to each neuronal discharge time a number chosen randomly between +/- $\Delta T/2$, where $\Delta T = T_{max} - T_{min}$. We will refer to this procedure as $\Delta T$ randomization. One may also choose to run the same data set through this procedure more than once, to improve the statistics of $\Delta T$ randomization.

## Results

The data set consists of neuronal unit discharges from one particular neuronal unit, recorded from the basal ganglia of a non-human primate. The number of discharges recorded from this unit is 1364, accumulated over a time span of 275 secs. The distribution of interspike intervals is well described by a single decaying exponential with a time constant of 0.2 sec (Fig 1). Thus, this process would seem to be Poissonian. Evaluation with FFT (Fig 2) and CS (Fig 3) show no structure in the frequency domain.

In contrast, applying the DOOD algorithm shows sharp peaks at 70 and 20 Hz, which persist when a small amount of friction is turned on and regardless of whether one applies $\Delta T$ randomization or not (Fig 4, a and b). We also looked at how *S(t)* changes in time for the 20 and 70 Hz oscillators. They appear to come and go with a timescale of 30 to 60 secs (Fig 4c).

**Discussion**

The DOOD algorithm is a conceptually simple algorithm for the detection of oscillatory behavior. We believe it may be most useful for the detection of transient oscillations of point processes such as neuronal unit discharges as well as extracellular local field recordings of population spikes. Further testing on model data is necessary. However, the peaks at 20 and 70 Hz are so striking in our preliminary investigations that we suspect they are real. Indeed, the 70 Hz oscillator in the basal ganglia is known in the literature as the "prokinetic" oscillator while the 20 Hz oscillator is known as the "anti-kinetic" oscillator, because of the roles these oscillators are suspected to play in various movement disorders [4, 5].

If the basal ganglia oscillators we have detected are real, then one wonders why FFT and CS failed to detect them. One reason may be that FFT and CS scale linearly with the signal to noise ratio, even when "on resonance" with a data oscillator. In contrast, a mathematical oscillator in the DOOD algorithm interacts in a supralinear way with a data oscillator. To see this, consider Eq (7) but neglect for now the friction term:

$$S = \dot{x}h \qquad \text{Eq (10)}$$

The factor of $h$ on the right hand side provides a linear response of $S$ to input data, but the velocity factor $\dot{x}$ also increases as $h$ increases. Even if a data oscillator appears, disappears and then appears again, the velocity factor provides a cumulative memory effect for that oscillator, as long as the friction is not too high. The net effect is a supralinear response.

To compare DOOD vs FFT and CS, we generated data by adding Poisson noise to a discrete oscillator. The model data oscillator was taken to have a frequency of 70.231 Hz (chosen to avoid a round number), and a probability of transmission of $p_o$; that is, at every cycle of the model data oscillator, the probability whether it actually fires or not is given by $p_o$. The Poisson noise has a time constant of 5.372 Hz, to model the experimental timescale from our basal ganglia data. The FFT data was analyzed using non-overlapping sliding time windows consisting of 4096 time points in increments of 1 msec. The power spectrum then has a frequency increment of approximately 0.244 Hz. The time windows are accumulated over 2000 firing events (oscillator plus Poisson noise), and averaged. The same frequency increment of 0.244 Hz was chosen for DOOD and CS analysis.

In Fig (5), the z-values of the 70 Hz peak as a function of $p_o$ are shown. The DOOD algorithm with $g_0 = 10^{-3}$ performs best, while CS is worst. It is helpful to adjust $g_0$ to see which peaks damp with friction and which peaks are retained. Too high a value of $g_0$ damps out all oscillations. In Fig (6) are shown representative spectra with the three methods with $p_o = 0.2$. Correct interpretation of all three spectra requires understanding that harmonics appear as multiples of the fundamental frequency. These harmonics are damped using the DOOD algorithm, are less damped using the FFT algorithm, and can be difficult to interpret using the CS algorithm.

In summary, we describe a simple new algorithm for detecting oscillatory behavior in discrete data. In certain circumstances, it can be more sensitive to oscillatory behavior than either the fast Fourier or circular statistics techniques. Caution in interpretating spectra is still warranted and comparison of spectra with multiple methods remains advisable.

**Acknowledgements:** DH is supported by National Institutes of Health K12 Roadmap Project number 8K12RR023268-03. EBM is supported in part by an American Parkinson Disease Association (APDA) Advanced Center for Research Grant, the Roger Duvoisin Fellowship of the APDA, and NIH grant 5P51 RR000167 to the National Primate Research Center at the University of Wisconsin – Madison.

Fig 1

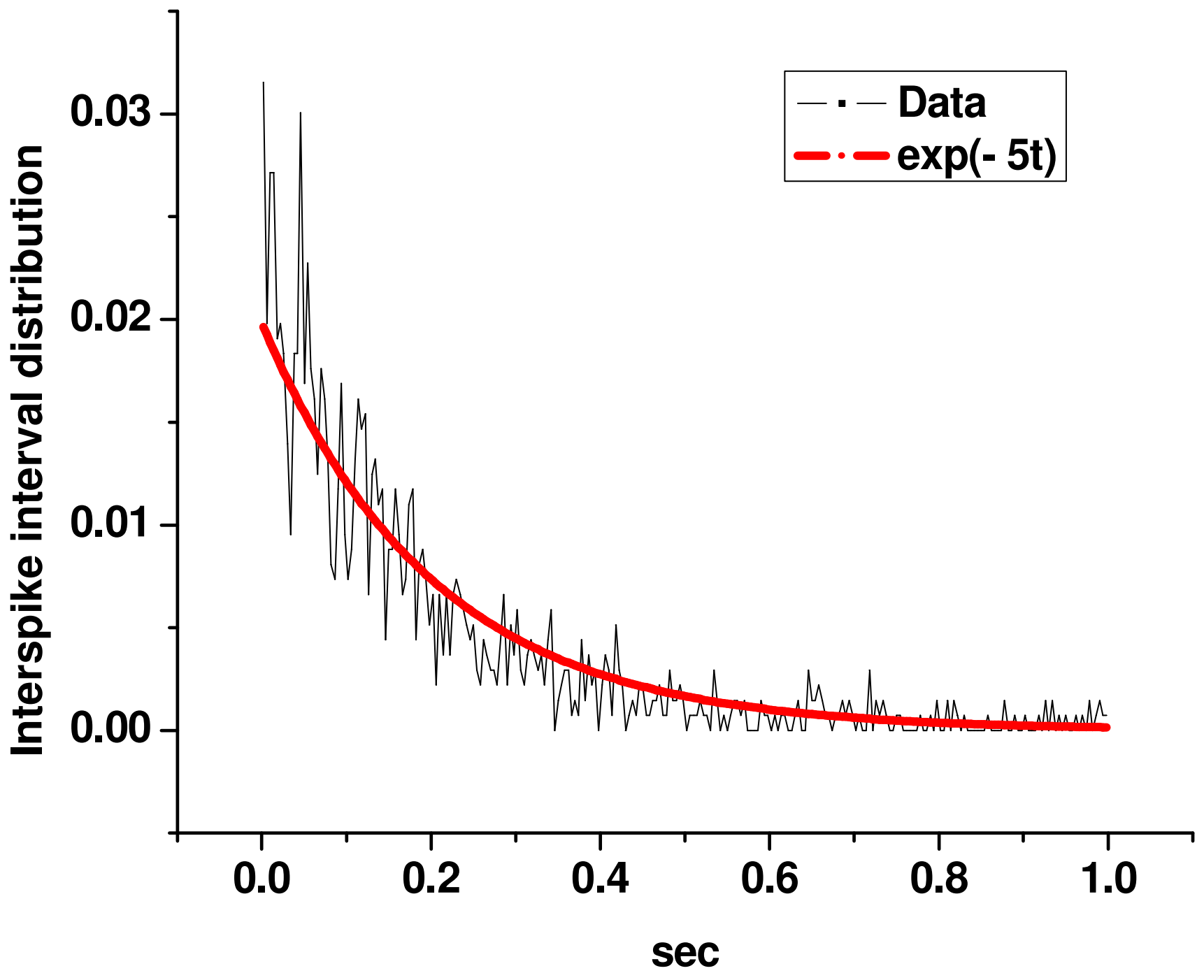

Fig 1 caption: Black line: experimental interspike intervals. Red line: single exponential fit, with a time constant of 0.2 secs (5 Hz).

Fig 2

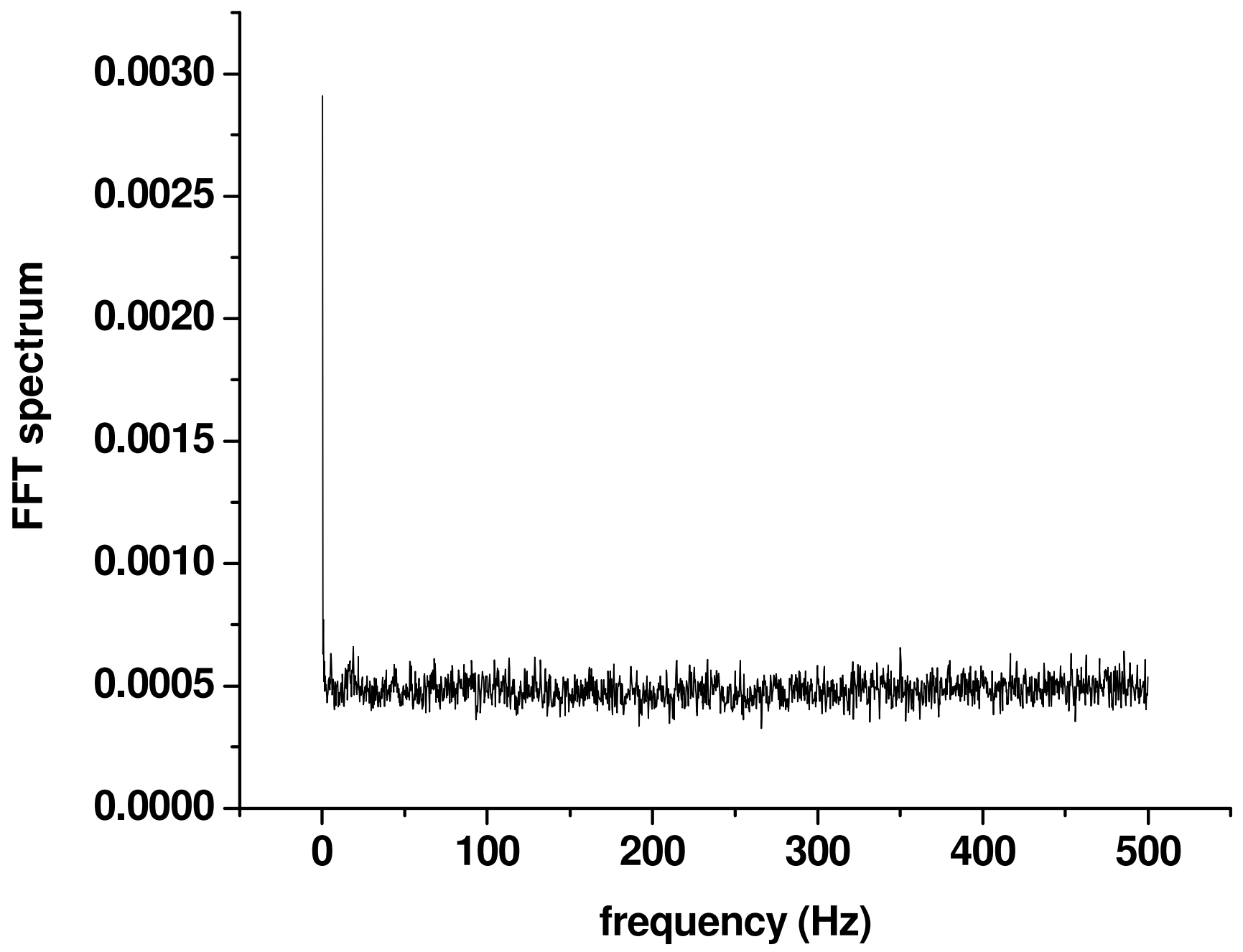

Fig 2 caption: FFT of data using non-overlapping sliding time windows. In this example, each time window consists of 4096 time bins, each 1 msec wide. The FFT spectrum from each time window is accumulated and then averaged over the entire data set. Varying the time bin width up to 10 msecs and increasing the number of bins per time window up to 16,384 produced no qualitative difference. That is, no structure is evident no matter how the FFT time windows were chosen.

Fig 3

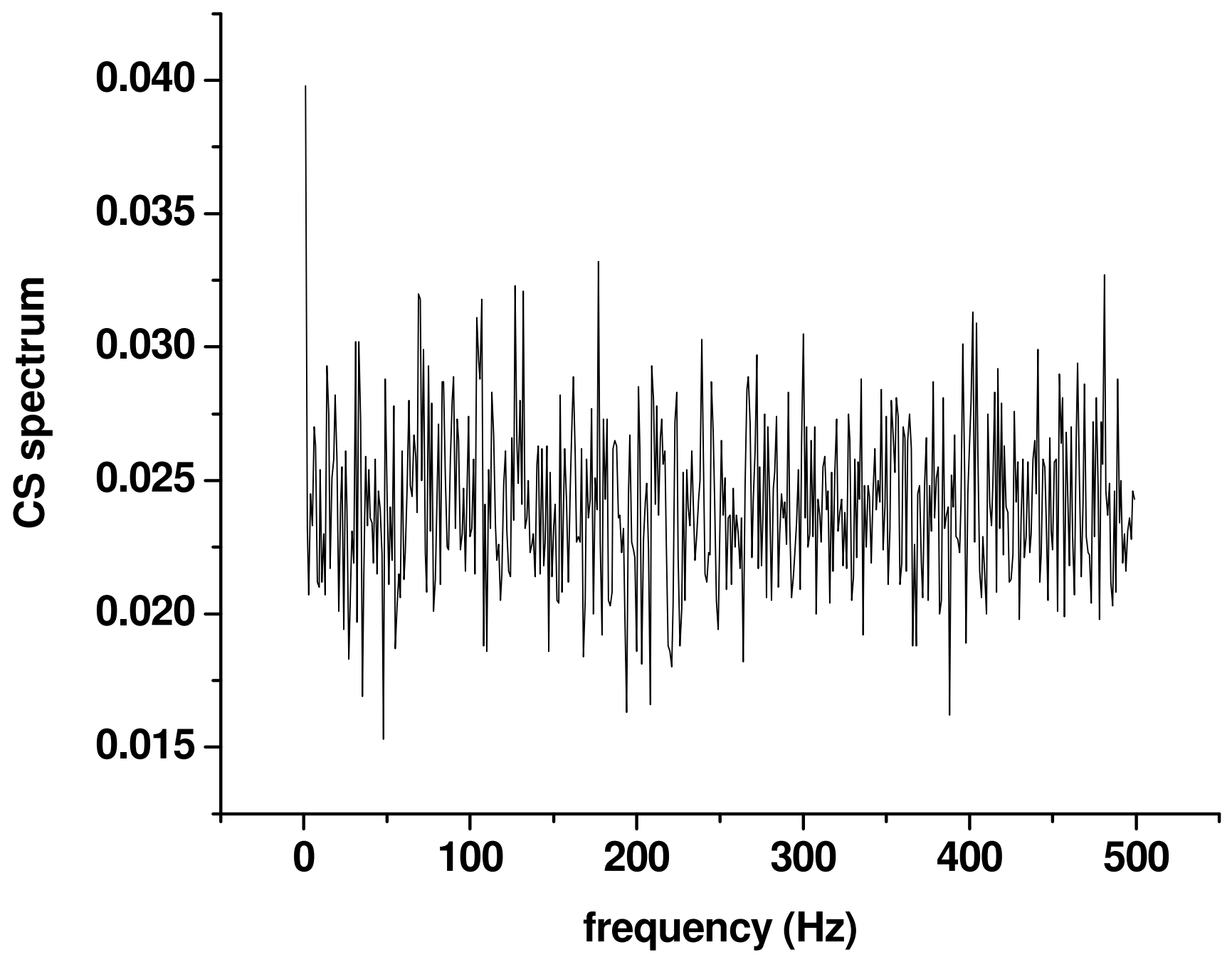

Fig 3 caption: Circular statistics spectrum using cycles spaced at 1 Hz intervals. Changing the spacing down to 0.1 Hz and up to 2 Hz did not change results qualitatively. That is, no structure is evident.

Fig 4a.

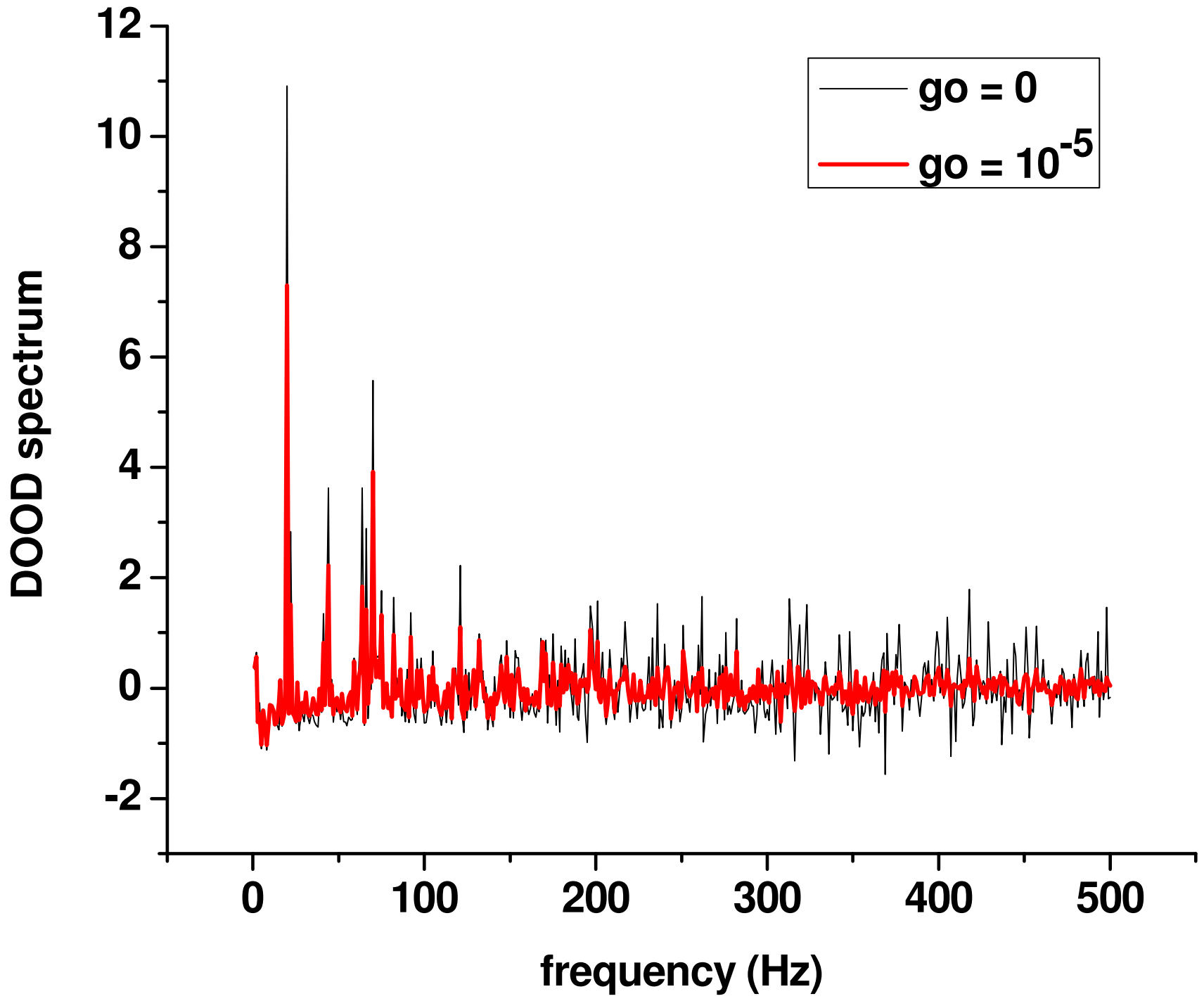

Fig 4a caption: DOOD spectrum with Δf = 1 Hz, ΔT randomization turned on, and number of repetitions at 10 (that is, the entire data set is run through the ΔT randomization process 10 times). Black line: no friction. Red line: friction with $g_0 = 10^{-5}$. The two largest peaks are at 20 and 70 Hz. The y-axis is expressed in Z-values.

Fig 4b

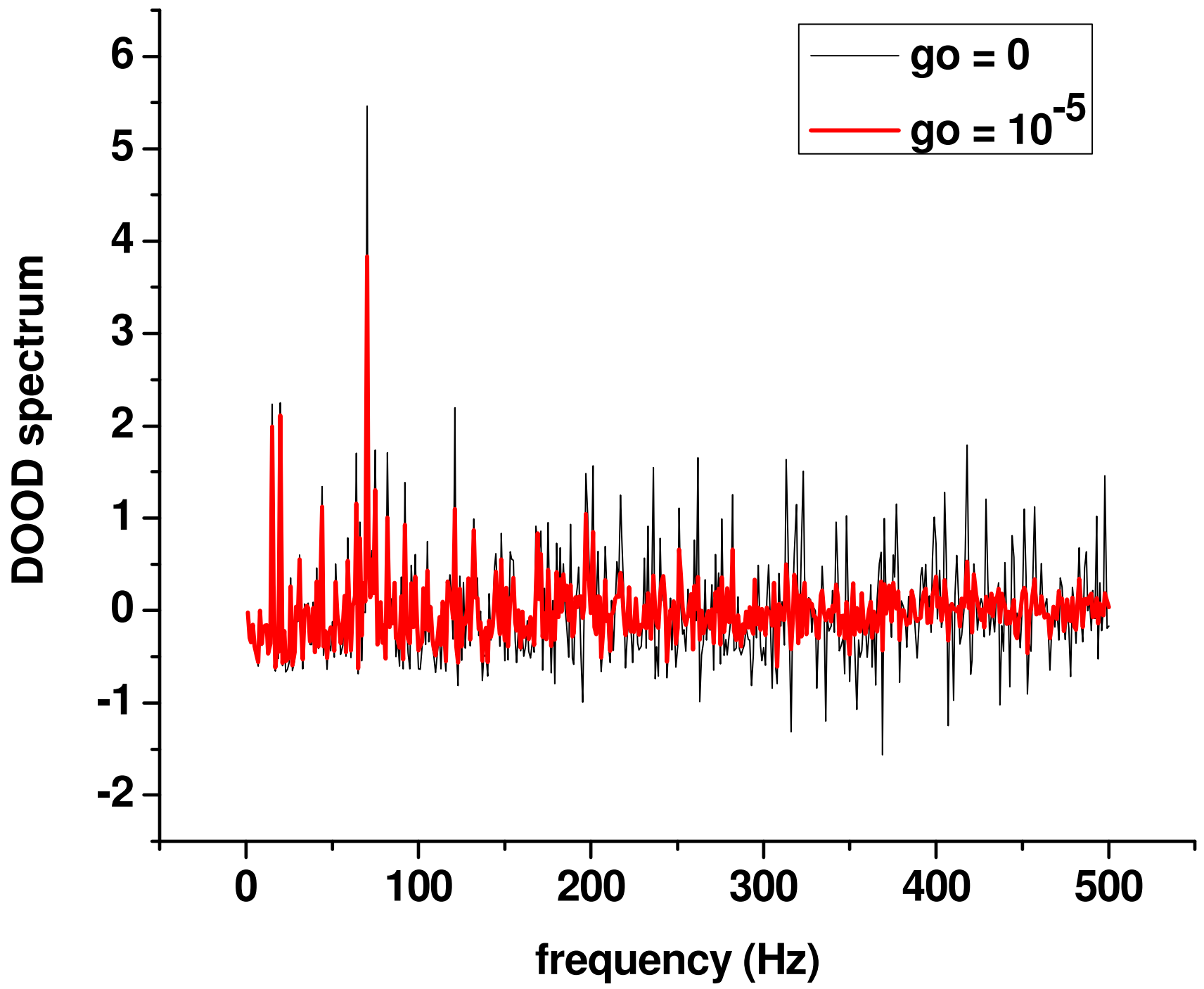

Fig 4b caption: Same as Fig 4a except that ΔT randomization is turned off. In this case, the largest peak occurs at 70 Hz, the next largest at 20 Hz.

Fig 4c

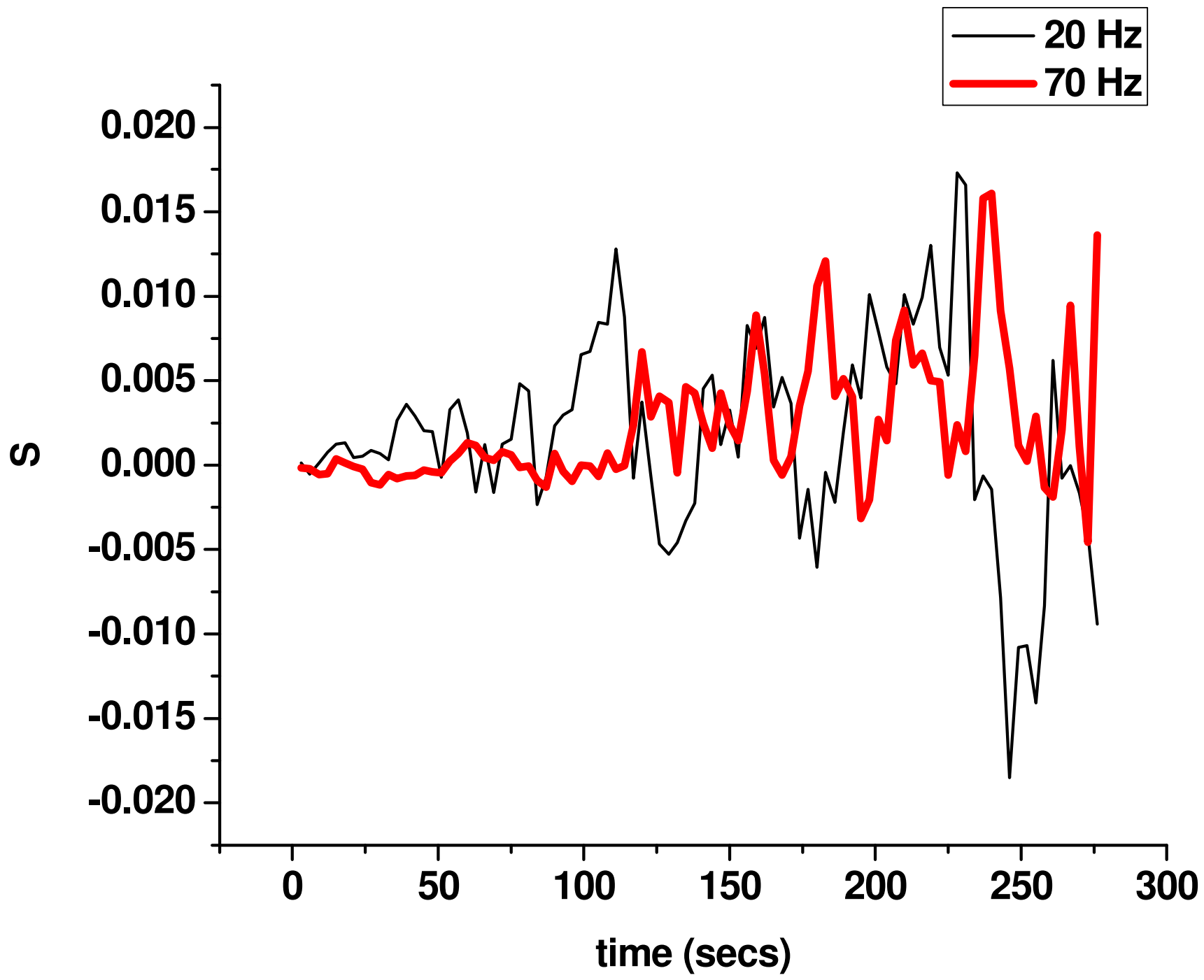

Fig 4c caption: S as a function of time for the 20 and 70 Hz oscillators. Friction is taken as $g_0 = 10^{-5}$, and averaging time is $\tau_0 = 3$ secs. Thin black line: 20 Hz oscillator. Thick red line: 70 Hz oscillator.

Fig 5

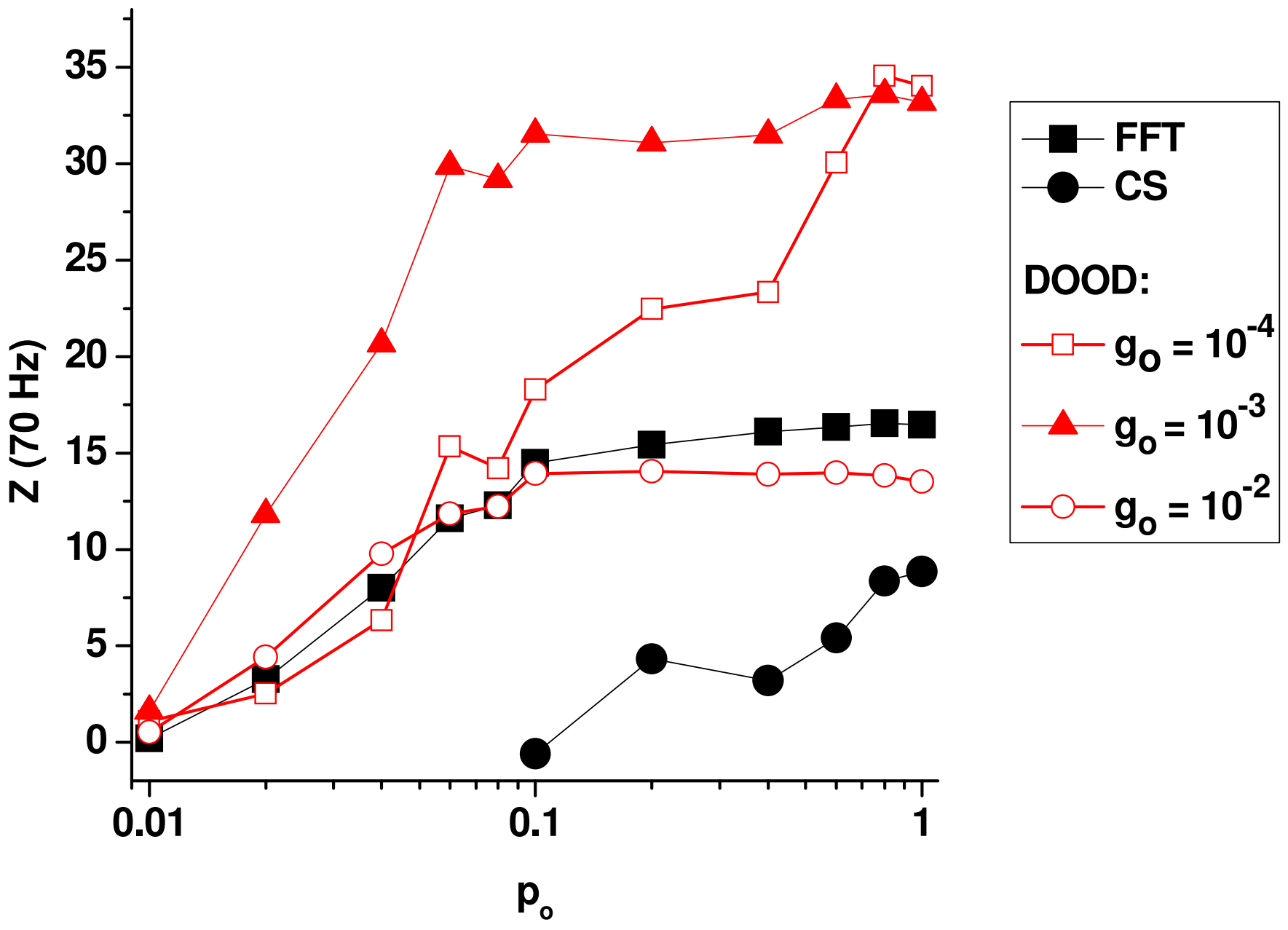

Fig 5 caption: The Z values of the 70 Hz peak is shown as a function of the transmission probability $p_o$. The x-axis is on a logarithmic scale. The DOOD algorithm performs best with a friction constant of $g_o = 10^{-3}$. The CS algorithm performs most poorly.

Fig 6a

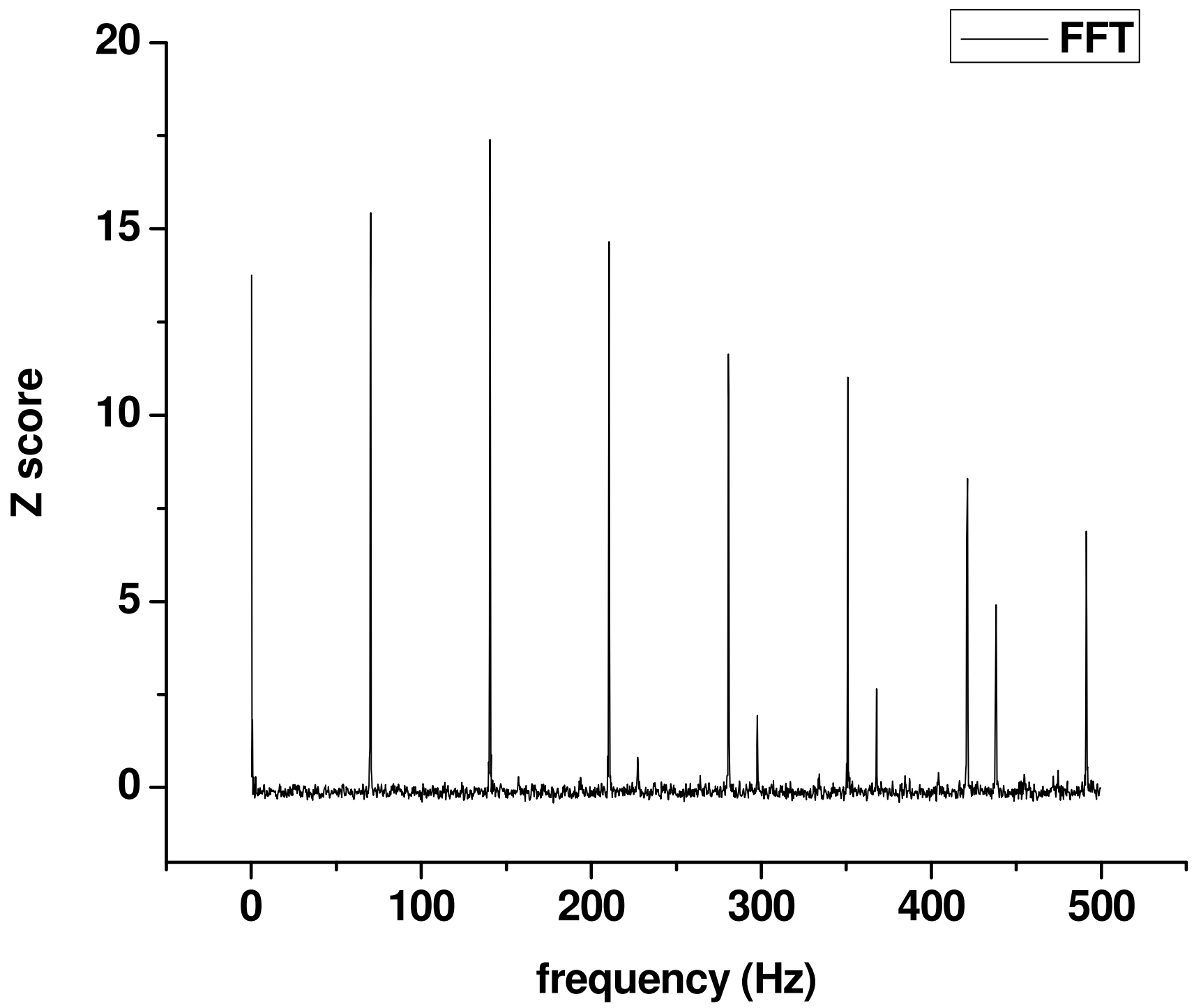

Fig 6a caption: Sample FFT spectrum in terms of Z score for computer generated data consisting of a discrete oscillator of frequency circa 70 Hz, transmission probability of 0.2, Poisson noise with time constant of circa 5 Hz. Note fundamental frequency at 70 Hz, plus higher harmonics.

Fig 6b

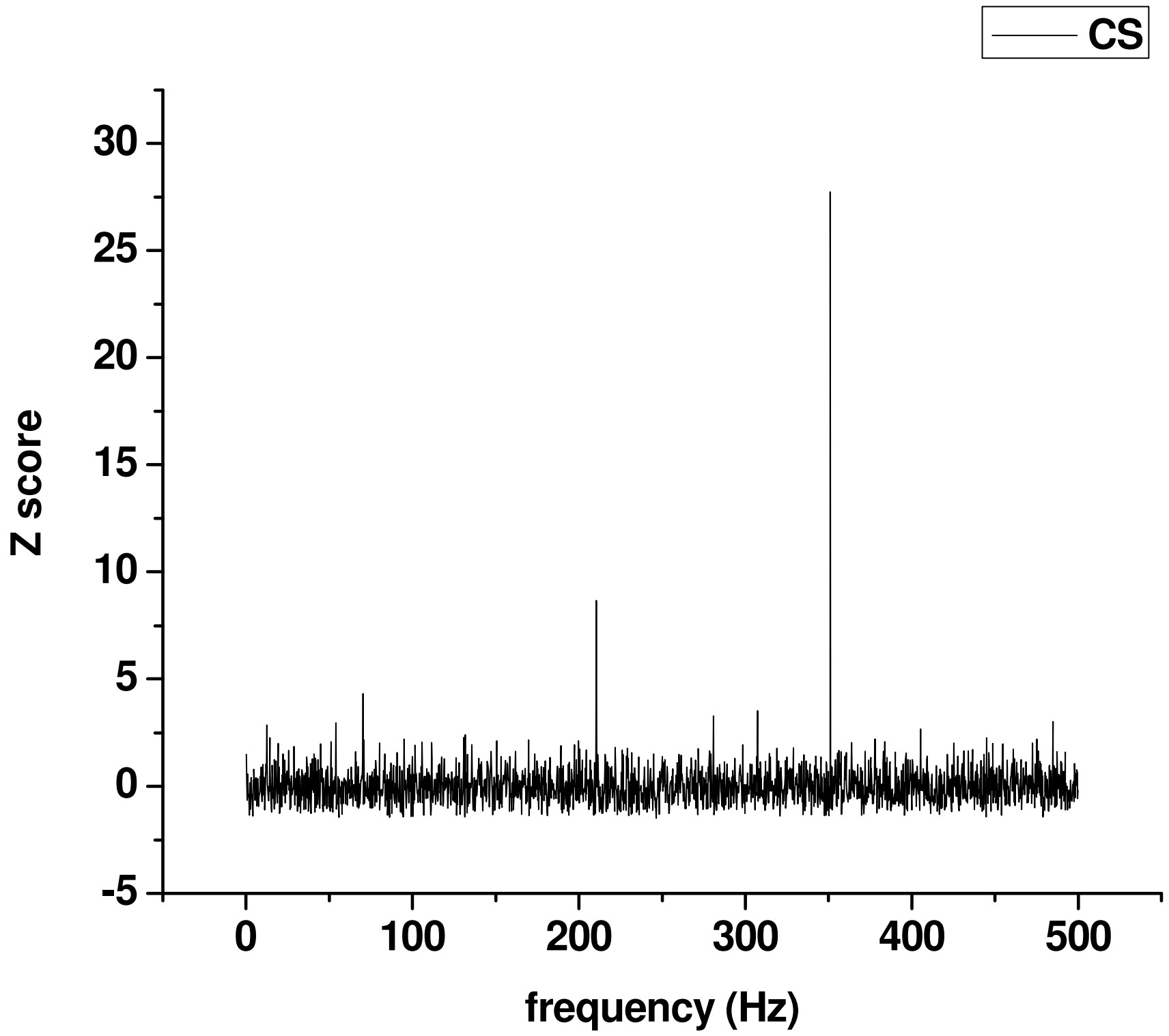

Fig 6b caption: Sample CS spectrum in terms of Z score for computer generated data consisting of a discrete oscillator of frequency circa 70 Hz, transmission probability of 0.2, Poisson noise with time constant of circa 5 Hz. Note fundamental frequency at 70 Hz, plus higher harmonics at 210 Hz and 350 Hz. The other harmonics are not as clearly discerned, making this spectrum rather misleading.

Fig 6c

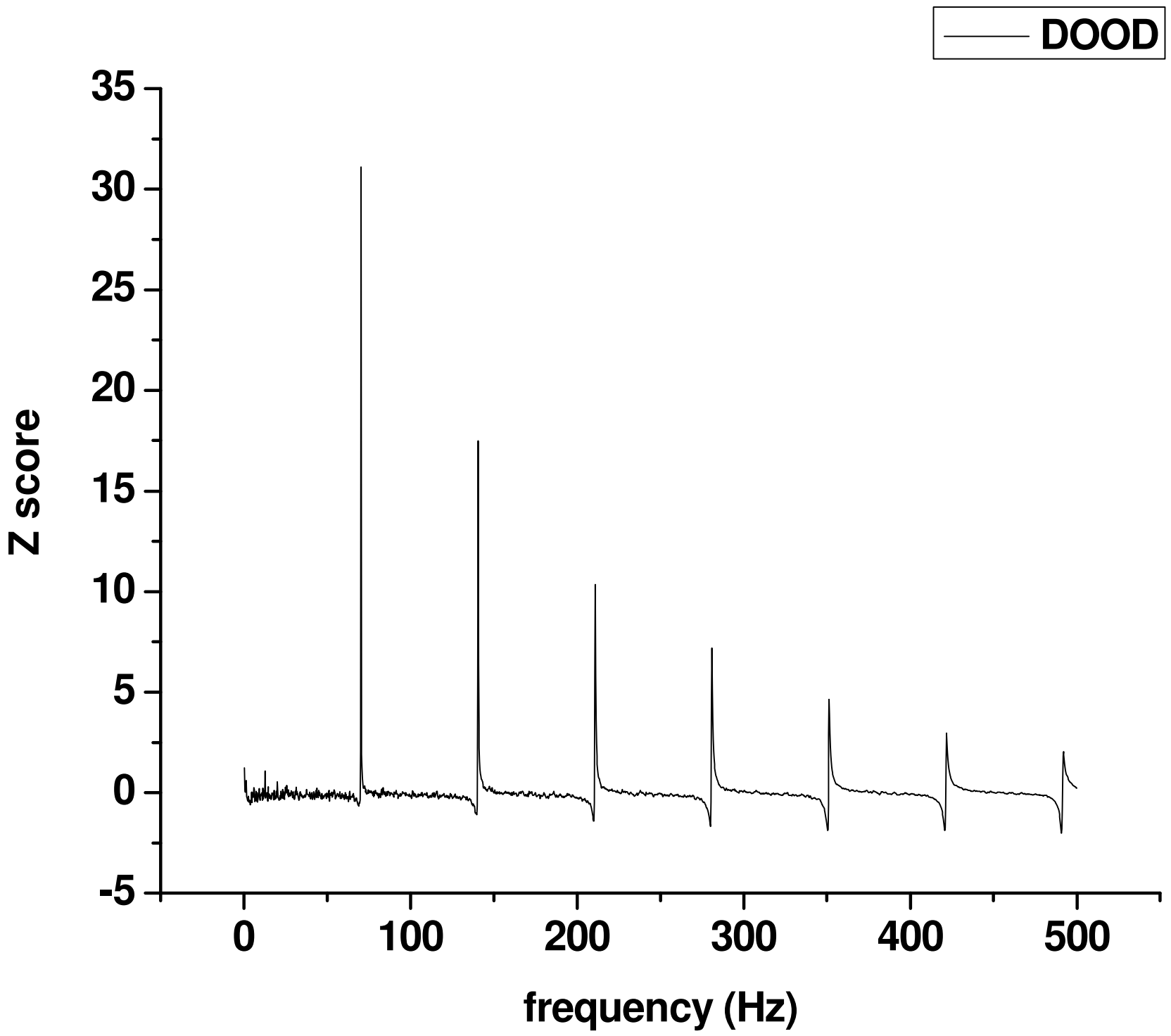

Fig 6c caption: Sample DOOD spectrum in terms of Z score for computer generated data consisting of a discrete oscillator of frequency circa 70 Hz, transmission probability of 0.2, Poisson noise with time constant of circa 5 Hz. Note fundamental frequency at 70 Hz, plus all higher harmonics. The higher harmonics are clearly damped, making this spectrum easily interpretable.